\input harvmac

\def\p{\partial}

\def\half{{1\over 2}}

\Title{hep-th/0212345}{\vbox{\centerline{PP-wave Black holes and The Matrix 
Model}}}
\vskip20pt

\centerline{Miao Li}
\vskip 10pt
\centerline{\it Institute of Theoretical Physics}
\centerline{\it Academia Sinica}
\centerline{\it Beijing 100080} 
\centerline { and}
\centerline{\it Department of Physics}
\centerline{\it National Taiwan University}
\centerline{\it Taipei 106, Taiwan}
\centerline{\tt mli@phys.ntu.edu.tw}

\bigskip

We discuss the sizes of a black hole in the M theory pp-wave background,
and how the transverse size can be reproduced in the matrix model.

%\draft
\Date{Dec. 2002}

\nref\bmn{D. Berenstein, J. Maldacena and H. Nastase, ``Strings in flat 
space and pp waves from ${\cal N}=4$ Super Yang Mills," hep-th/0202021;
R. Penrose, ``Any spacetime has a plane wave as a limit," in 
Differential Geometry and Relativity, Reidel, Dordrecht, 1976;
J. Figueroa-O'Farrill and G. Papadopoulos, ``Homogeneous fluxes, branes 
and a maximally supersymmetric solution of M-theory," JHEP 0108 (2001)
036,  hep-th/0105308;
M. Blau, J. Figueroa-O'Farrill, C. Hull and G. Papadopoulos, ``Penrose 
limits and maximal supersymmetry," hep-th/0201081;
 R. R. Metsaev, ``Type IIB Green-Schwarz superstring in plane 
wave Ramond-Ramond background," Nucl.Phys. B625 (2002) 70-96,
hep-th/0112044.}
\nref\mli{M. Li, ``Correspondence Principle in a PP-wave Background,"
hep-th/0205043, Nucl.Phys. B638 (2002) 155-164.}
\nref\ppmt{K. Dasgupta, M. M. Sheikh-Jabbari, and M. Van Raamsdonk,
``Matrix Perturbation Theory For M-theory On a PP-Wave,"
hep-th/0205185, JHEP 0205 (2002) 056;
``Protected Multiplets of M-Theory on a Plane Wave,"  hep-th/0207050,
JHEP 0209 (2002) 021;
N. Kim and J.-H. Park, ``Superalgebra for M-theory on a pp-wave,"
hep-th/0207061, Phys.Rev. D66 (2002) 106007;
N. Kim, Jan Plefka, ``On the Spectrum of PP-Wave Matrix Theory,"
hep-th/0207034;
N. Kim, K. M. Lee and Piljin Yi, ``Deformed Matrix Theories with N=8 
and Fivebranes in the PP Wave Background," hep-th/0207264,
JHEP 0211 (2002) 009;
J. Maldacena, M. M. Sheikh-Jabbari and M Van Raamsdonk,
``Transverse Fivebranes in Matrix Theory," hep-th/0211139.}
\nref\mbh{T. Banks, W. Fischler, I. Klebanov and L. Susskind,
``Schwarzschild black holes from matrix theory I and II", 
hep-th/9709091, hep-th/9711005; I. Klebanov and L. Susskind, ``
Schwarzschild black holes in various dimensions from matrix theory", 
hep-th/9709108; G. Horowitz and E. Martinec, ``Comments on black 
holes in matrix theory", hep-th/9710217;
M. Li, ``Matrix Schwarzschild black holes in large N
limit", hep-th/9710226, JHEP 9801 (1998) 009;
M. Li and E. Martinec, ``Probing matrix black holes," hep-th/9801070.}
\nref\samir{S. D. Mathur, A. Saxena, Y. K. Srivastava,
``Scalar propagator in the pp-wave geometry obtained from 
$AdS_5\times S^5$," hep-th/0205136, Nucl.Phys. B640 (2002) 367-378.}
\nref\dkps{ M.R. Douglas, D. Kabat, P. Pouliot and S. H. Shenker,
``D-branes and Short Distances in String Theory," hep-th/9608024,
 Nucl.Phys. B485 (1997) 85-127;
K. Becker, M. Becker, J. Polchinski and A. Tseytlin
Phys. Rev. D56 (1997) 3174-3178, hep-th/9706072.}

PP-wave backgrounds, as a limit of AdS and some inner spheres, have proven 
an interesting place to test ideas of holography and the correspondence between
string/M theory in such a background and some supersymmetric gauge theory
\bmn. In a previous paper \mli, we discussed possible correspondence between
string states and black hole states in such a background, when the string coupling
is fined tuned for a given total oscillation number of the strings.
As we shall see, the estimate of the transverse gravitational size of a 
black hole in that paper is correct, but the longitudinal size given there is incorrect.

In the original paper, the first paper in \bmn, a matrix model is given for 
the PP-wave background obtained from the M theory AdS backgrounds.
This matrix model was further discussed in \ppmt. Here we are interested
in the question of whether this matrix model can reproduce at least the
transverse size of a black hole for a given energy. As we shall argue,
indeed it can, provided this model can produce a certain form of interaction
between two partons. This form, we will argue, shall come out naturally
from a computation in the matrix model, although we will not undertake
such a calculation in this paper.

The spirit of the analysis in this paper is that of \mbh, where Schwarzschild
black holes in a flat spacetime is analyzed using matrix theory. The
main ingredients in that analysis are to assume that a black hole is a bound
state of partons which are completely virialized, and to assume that 
the minimal uncertainty relation is obeyed by partons.

We will start with an estimation of the transverse size of gravitating point
source in the pp-wave background, and also discuss its longitudinal size.
Then we turn to the matrix model, give a general analysis of the possible
loop corrections to the interaction potential between two partons. We argue
that the one loop correction must assume the form
\eqn\intpot{V_1=c{G\over R^3}{v^5\over r^8},}
where $c$ is a dimensionless numerical coefficient, $G$ is the Newton 
constant in 11 dimensions, $R$ is the longitudinal radius in the direction
$x^-$, $v$ is the relative velocity between two partons, and $r$ is their
transverse separation. Although the power in the velocity dependence
sounds a little strange, we will argue that this should be most important
term in the one-loop correction, if $\mu r<v$. Happily, as we shall see, this is the
right term to reproduce the correct transverse horizon size of a 
black hole. The more accurate form of \intpot\ appears in (30)
and breaks time reversal invariance. This is possible, because the existence
of the four form field strength indeed breaks time reversal invariance.
We leave the computation of the interaction between two partons
to a future work.

\noindent {\it The gravitational sizes}

The metric of the M theory pp-wave background is 
\eqn\metric{\eqalign{ds^2&=-4dx^+dx^--[({\mu^2\over 9}\sum_i
(x^i)^2+{\mu^2\over 36}\sum_a(x^a)^2](dx^+)^2+(dx^A)^2,\cr
i=&1,2,3,\quad a=4,\dots 9, \quad A=i,a.}}
Since $x^+$ is taken as our time in the light-cone approach, there
is a harmonic potential in each transverse direction, even for a
massless particle. This term is the source for a kind of dimensional
transmutation in the Newtonian potential of a point source.

To see the origin of the dimensional transmutation, let us start with
a flat spacetime, where we shall get the standard Newtonian potential.
To be general, let us work in $D$ dimensional spacetime, the
retarded Green's function in the momentum space is simply
$1/[(\omega+i\epsilon)^2-k^2)]$. It proves simplest to work in
the light-cone frame to carry out the Green's function in the coordinate
space, by first integrating over $k_+$, and later integrating over the
transverse momentum space, the result is
\eqn\fgreen{G(x)={1\over 4\pi^{D/2}}e^{-\pi Di/4}\theta(x^+)\int_0^\infty 
dk_-k_-^{{D-4\over 2}}[e^{-ik_-(x^2-i\epsilon)}-e^{ik_-(x^2+i\epsilon)}],}
where $x^+=x^+x^--x_{\perp}^2$, $x_\perp$ is the transverse
coordinates. We will drop the subscript $\perp$ in the following
expressions.

To estimate the gravitational sizes of a moving particle with longitudinal
momentum $p^\pm$, let us use the component of the stress
tensor $T_{+-}$ (see \mli)
\eqn\stresst{T_{+-}(x)=p^-\delta (x^--ax^+)\delta^{D-2}(x),}
where $a=p^-/p^+$, and we have been loose about the numerical 
coefficient, since we are interested only in the functional form of
the gravitational size. The metric perturbation $h_{+-}$ is then
\eqn\meper{h_{+-}=G\int G(x-y)T_{+-}(y)dy,}
again a numerical coefficient is not taken into account. Integration
over $y^-$ and $y$ is trivial. Let $\tau =x^+-y^+$, the metric
perturbation is
\eqn\mper{h_{+-}=I_+-I_-,}
with
\eqn\iper{I_+\sim Gp^-\int_0^\infty dk_-k_-^{{D-4\over 2}}\int_0^\infty
d\tau\exp\left(-ik_-[a\tau^2+\tau (x^--ax^+)-x^2-i\epsilon]\right),}
and $I_-$ is given by a similar formula, with $k_-,\epsilon
\rightarrow -k_-, -\epsilon$. The integration over $\tau$
can be carried out approximately, if $k_-a$ is sufficiently large.
However, if $x^--ax^+$ is positive, the arguments in the exponential
in \iper\ containing $\tau$ are all positive, the integral is damped
greatly. This simply implies that the particle at time $x^+$ has not
yet reached the longitudinal position $x^-$.If $x^--ax^+<0$, the
integral over $\tau$ can be carried out approximately, we have
\eqn\appi{I_+\sim Gp^-/\sqrt{a}\int_0^\infty  dk_-k_-^{{D-5
\over 2}}\exp\left(ik_-[x^2+{(x^--ax^+)^2\over 4a}+i\epsilon]\right).}
Remember that the integral over $\tau$ results in a factor
$1/\sqrt{ak_-}$ which is independent of $x$. Finally, we carry out
the integral over $k$ by rotating the contour. Put $I_+$ and $I_-$
together,
\eqn\poper{h_{+-}\sim G(p^-/\sqrt{a}) [x^2+{(x^--ax^+)^2\over 4a}]
^{-{D-3\over 2}}.}

Note that in \poper, $p^-/\sqrt{a}=\sqrt{p^+p^-}=m$, just the invariant
mass of the boosted particle, this must be the case since the strength
of the gravitational field must be proportional to the invariant mass
in a flat spacetime, due to Lorentz invariance. When we look at near
the trajectory of the particle in the longitudinal direction where
$x^--ax^+=0$, the gravitational potential assumes the form
\eqn\lpoten{h_{+-}\sim Gmr^{-(D-3)},}
where $r=|x|$. The above is the correct Newtonian potential, for the
dependence on the transverse direction should be boost invariant.
Thus, the gravitational size in the transverse direction is just
$r_0^{D-3}\sim Gm$. Set $r=0$ in \poper, we obtain the gravitational
size in the longitudinal direction $\Delta x^-=\sqrt{a}r_0$. Let
$p^\pm =e^{\pm \alpha}m$, $\alpha$ is the boost parameter, we
find $\Delta x^-=e^{-\alpha}r_0$, namely, there is the standard Lorentz
contraction in the longitudinal direction.

Having discussed the estimate of gravitational sizes of a boosted particle
in a flat spacetime, we are ready to discuss these in a pp-wave
background. For definitiveness, we work in 11 dimensions. The 
exact scalar Green's function can be computed in a similar fashion as
in \samir, we will do it in the appendix. Here for simplicity, we
will work with a metric with $SO(9)$ symmetry, namely with a 
fictitious metric
\eqn\ficmetric{ds^2=-4dx^+dx^--(\mu^2\sum_a(x^a)^2)(dx^+)^2+
(dx^a)^2,}
where $a$ runs from 1 to 9. To simplify all the formulas in the following,
we do the rescaling $x^\mp\rightarrow \mu^\pm x^\mp$. We will
recover the factor $\mu$ in the final result. Although the above
metric is quite different from \metric, the physics captured by it
will not be so different. The sizes in two sets of transverse directions
in the real space \metric\ may be different, but their dependence on
physical parameters such $p^\pm$ and $\mu$ will be similar.

The Green's function in the background \ficmetric\ can be computed
along the line in \samir, and is given by a similar formula as in
\fgreen. In fact, we simply replace $(x-y)^2$ in that formula by $\Phi$
which is
\eqn\ppdis{\Phi =-4(x^--y^-)\sin (x^+-y^+)-[1-\cos(x^+-y^+)](x^2+y^2)
+(x-y)^2,}
where $x,y$ are the transverse locations of two spacetime points.
$\Phi$ maybe interpreted as the invariant spacetime distance between
two points in the pp-wave background.

The component of stress tensor $T_{+-}$ is still given by \stresst, we
plug it as well as the Green's function \ppdis\ into \meper\ and obtain
\eqn\pppot{h_{+-}\sim Gp^-\int dk_-k_-^{{D-4\over 2}}\int_0^\infty d\tau
e^{ik_-(\Phi(\tau)+i\epsilon)}+\dots,}
where the first term is similar to $I_+$ in \mper, and $\dots$ denotes the
term similar to $I_-$, $\Phi(\tau)$ is given by
\eqn\integ{\Phi(\tau)=-4(x^--ax^++a\tau)\sin\tau +x^2\cos\tau.}
Thus, the integral over $\tau$ in \pppot\ is more involved. We will use the
stationary method to approximate this integral. The stationary point 
satisfies
\eqn\stapt{\Phi'(\tau)=(x^2+4a)\sin\tau +4(x^--ax^++a\tau )\cos\tau =0.}
When $x^--ax^+=0$, on the light-cone trajectory, one stationary point is
clear, it is just $\tau =0$. There is another stationary point, it contributes
much less than does the point $\tau =0$. Near $\tau=0$, $\Phi(\tau)$ expanded to
the second order is 
\eqn\exppho{\Phi(\tau)=x^2-(4a+{x^2\over 2})\tau^2.}
Note that the term proportional to $\tau^2$ is quite different from that 
in \iper, now it depends on $x^2$ as well as on $a$. When $a\gg x^2$,
we go back to \iper\ and obtain a result same as in the flat case. If 
$x^2\gg a$, the result is completely different. Integrating out $\tau$
as well as $k_-$, we have
\eqn\pphm{h_{+-}\sim Gp^- |x|^{-(D-3)}(4a+{x^2\over 2})^{-1/2},}
apparently, if $a\gg x^2$, we obtain the Newtonian potential in the
flat space. Rescaling back $p^\pm\rightarrow \mu^{\pm 1} p^\pm$, the
condition is $a\gg \mu^2x^2$, this is just to say that the harmonic
potential in the metric is not important. For $a\ll \mu^2x^2$, the result
is
\eqn\mainr{h_{+-}\sim {Gp^-\over \mu |x|^{D-2}},}
this is a result already obtained in \mli. The Newtonian potential is modified
by a factor $1/(\mu |x|)$. For small $x$, the result of \mli\ is incorrect,
the reason is that one cannot truncate the Green's function to the first mode
in this case as done in that paper.

Thus, the transverse gravitational size is given by
\eqn\transs{r_0^{D-2}={Gp^-\over \mu}.}
For $D=11$, the case of interest, the exponent is $9$ rather than $8$.
Also, the transverse size no longer depends on the invariant mass, it
depends both on the light-cone energy as well as on the parameter $\mu$.
We want to note that although the metric \metric\ and that in \ficmetric\
are not Lorentz boost invariant, there is a generalized Lorentz boost
invariance
\eqn\boostt{\eqalign{x^\pm &\rightarrow e^{\pm\alpha} x^\pm ,\quad
p^\pm\rightarrow e^{\pm\alpha}p^\pm ,\cr
\mu &\rightarrow e^{-\alpha}\mu.}}
Formula \transs\ is invariant under this generalized boost.

The other extreme points to look at are when $x^2=0$, we want to know
the longitudinal size of the gravitating point along the origin in
the transverse space. We need to solve the stationary point
\eqn\lstan{a\sin\tau +(x^--ax^++a\tau)\cos\tau=0.}
Let the solution be $\tau_0>0$. Near this point
\eqn\expphi{\Phi(\tau)=-2a{1+\cos^2\tau_0\over \cos\tau_0}\delta\tau^2
+4a{\sin^2\tau_0\over\cos\tau_0}.}
Using the stationary method, we find
\eqn\lpot{h_{+-}\sim Gp^-a^{-{D-2\over 2}}|\sin\tau_0|^{-(D-3)}
|\cos\tau_0|^{{D-2\over 2}}(1+\cos^2\tau_0)^{-1/2}.}
Apparently, the above result is periodic in $x^--ax^+$.
If $\tau_0(x^--ax^+)$ is a solution to \lstan, then 
$\tau_0(x^--ax^+-2\pi na)=\tau_0(x^--ax^+)+2\pi n$, and the
potential \lpot\ is not changed at all.
The exact result cannot be exactly periodic in $x^--ax^+$, since
although the function \integ\ is periodic under the double shifts
$x^--ax^+\rightarrow x^--ax^+-2\pi na$, $\tau\rightarrow \tau
+2\pi n$, but the range of the integration of $\tau$ is also changed.
Thus, the gravitational tail in the longitudinal direction should
slowly die away.

\noindent {\it The matrix model analysis}

The matrix model proposed in the first reference of \bmn\ is described
by the action
\eqn\matrixa{\eqalign{S&=S_0+S_m,\cr
S_0&=\int dt\tr\left({1\over 2R}(D_tX^A)^2+{R\over 4l_p^6}[X^A,X^B]^2
+\psi D_t\psi +{iR\over l_p^3}\psi \gamma^A[\psi,X^A]\right),\cr
S_m&=\int dt\tr\left({1\over 2R}(-{\mu^2\over 9}(X^i)^2-{\mu^2\over 36}
(X^a)^2)-{\mu\over 4}\psi \gamma_{123}\psi -{\mu i\over 3l_p^3}
\epsilon_{ijk}X^iX^jX^k\right).}}
Again, the matrix model has the global invariance $SO(3)\times SO(6)$,
not $SO(9)$. For simplicity, in the following dimensional analysis, we
assume that the global invariance is $SO(9)$, it is not hard to 
amend our analysis for the action \matrixa, although for our purpose
it is not necessary to do this.

To perform a dimensional analysis, let us rescale $X$ and $\psi$ in
the following way
\eqn\resfs{X\rightarrow {l_p^3\over R}X,\quad \psi\rightarrow {l_p^3\over
R^{3/2}}\psi,}
The action after this rescaling takes the form
\eqn\resa{S={1\over g^2}[S_0+S_m],\quad g^2={R^3\over l_p^6},}
where the original matrix action $S_0$ now is independent of $l_p$ and
$R$, and the mass term $S_m$ is also independent of these parameters,
and depends only on $\mu$.

It is still possible to classify all the loop corrections into loops weighted by
the coupling constant $g^2$, although the new term $S_m$ does contain
an interaction vertex. Consider the $2\times 2$ matrices, describing a
system of two partons. We then formally expand the quantum effective
action in powers of $g^2$: $S_{eff}=\sum_n g^{2(n-1)}S_n$. We are interested
in the potential, and the potential has a similar expansion:
$V_{eff}=\sum_n V_n=\sum_n g^{2(n-1)}U_n$. Since the dimension of the 
coupling constant $g^2$ is $L^{-3}$, the dimension of $U_n$ is $L^{3n-4}$.
The rescaled $X$ or $r$, the separation of the two partons, has a dimension
$L^{-1}$, we can write $U_n=r^{4-3n}f_n$, now $f_n$ is a dimensionless
function. If we are interested in the bosonic part only, the dimensionless
function $f_n$ is a function of three sets of dimensionless quantities,
they are $x^i/r$, $v^i/r^2$ and $\mu/r$ (now since we are assuming
$SO(9)$ symmetry, $i$ runs from 1 to 9). Rescaling back to the original
coordinates, we have, in general
\eqn\nloop{V_n=Rl_p^{3n-6}r^{4-3n}f_n({x^i\over r},{l_p^3v^i\over Rr^2},
{l_p^3\mu\over Rr}).}
Of course, the above form is too general to be useful.

To fix the most important term at the one-loop level, without committing
concrete calculation starting with action \matrixa, we need to make some guess
or reasonable physics argument. Here is our reasonable argument.
Without the mass term, it is well known that the one-loop bosonic
interaction assumes the form ${v^4\over r^7}$, $v$ is the relative velocity
between the two partons \dkps. There is an additional dimensionful
coefficient $l_p^9/R^3$, of which $l_p^9/R^2$ can be interpreted as
$Gp^+(1)p^+(2)$. the combination $1/(Rr^7)$ can be interpreted as
the smearing of the Newtonian potential $1/r^8$ over the longitudinal
direction (produced by an infinity array of mirror images). Alternatively,
this potential can be obtained by considering a D0-brane moving in
the Aichelburg-Sexl background. As we already shown, in the pp-wave
background, the Newtonian potential for large $\mu r$ is no longer
$1/r^8$, but modified to $1/(\mu r^9)$. We expect that the D0-brane
interaction should be the smearing of this new potential in the longitudinal
direction, thus we should have a factor $1/(R\mu r^8)$ in the potential.
It must be also proportional to the Newton constant $G$ and $p^+(1)p^+(2)
=1/R^2$, thus in general we must have
\eqn\onep{V_1={G\over R^3\mu r^8}f(v^2, {v^ix^i\over r}).}
$f$ is a function of velocity and velocity components.
(In the flat background, the loop expansion was analyzed in the second
reference of \dkps.

Is it possible to reproduce the form \onep\ using \nloop\ at the one-loop
level? A general term of \nloop\ for $n=1$ is
\eqn\onelg{Rl_p^{-3}r[{l_p^3v^ix^i\over Rr^3}]^m[{l_p^6v^2\over R^2r^4}]^n
[{l_p^3\mu\over Rr}]^p.}
From the dependence of \onep\ on $\mu$, we determine $p=-1$.
Now \onelg\ becomes
\eqn\onelgg{{l_p^{3m+6n-6}\over \mu R^{m+2n-2}r^{2m+4n-2}}
v^{2n}[{v\cdot x\over r}]^m.}
If we demand the exponent of $l_p$ is $9$, namely $m+2n=5$,
we automatically have the exponent of $R$ in the denominator to
be 3 and the exponent of $r$ to be 8. Of course, on the dimensional
ground, only one of these two exponents is a free parameter, but
it is now determined by our general analysis. This analysis does not
help us to determine $m$ and $n$ separately, but for our purpose
we do not have to know these separately, so we will collectively denote
the one-loop potential by
\eqn\fonel{V_1={Gv^5\over R^3\mu r^8}.}
The fact that the power of $v$ is higher than that in the flat spacetime
may have something to do with our previous result on the modified
Newtonian potential, which is proportional to $p^-$ rather than the 
invariant mass.

Note that, for small $\mu r$ and very small  $v$, where $r$ is the 
characteristic separation
between two partons and their distance from the original, the original 
$v^4$ interaction is still most important. For this interaction to be smaller
than our new interaction \fonel, the condition is $\mu r<v$, namely the
relative velocity cannot be too small.
 
Of course, our general analysis can not replace a direct computation
of the one-loop potential in the matrix model, since it is highly nontrivial
for possible low order terms to cancel, to yield a result proportional
to $1/r^8$.

We now turn to a simple analysis of the ground bound states along 
the lines of \mbh. We assume that a simple black hole is composed
of $N$ partons in which each individual parton is essentially a 
distinguished constituent, and thus satisfies the minimal uncertainty
relation
\eqn\uncer{{1\over R}vr_0\sim 1,}
or 
\eqn\vel{v\sim {R\over r_0}.}
For the matrix model to be effective, $v$ must be small, thus $r_0$
is much greater than the longitudinal cut-off $R$. 

Next, each parton is subject to interaction with all partons of number
$N$ (for large enough $N$), and the total potential is roughly
\eqn\totp{NV_1={NGv^5\over R^3\mu r_0^8}.}
It must be the same order of the kinetic energy $v^2/R$. Using
the result \vel\ in $NV_1\sim v^2/R$, we find
\eqn\transd{r_0^{11}\sim {NGR\over \mu},}
it is certainly invariant under the generalized boost \boostt,
or $(\mu , R)\rightarrow e^{-\alpha}(\mu, R)$.
The above relation can be rewritten as 
\eqn\btrans{r_0^9\sim {G\over \mu}{N\over R}[{R\over r_0}]^2
\sim {Gp^-\over \mu},}
exactly the same formula as \transs, if we take $D=11$ in that formula.
We used I used $P^-\sim Nv^2/R$ and $v\sim R/r_0$ in arriving at
\btrans.

We see that indeed the matrix interaction can reproduce the formula
for the transverse horizon size of a black hole. In the flat background,
$N$ is taken as the entropy of the black hole \mbh, and there is
a general relation $S\sim N\sim r_0M$, $M$ is the invariant mass
of the black hole. In the pp-wave background, there is no boost
invariance, so we do not hope in general that the entropy is
boost invariant. However, we do have a generalized boost invariance
\boostt, and we can define the boost invariant mass by
$M=\sqrt{p^+p^-}\sim {N\over R}v$. Using the uncertainty relation
$v\sim R/r_0$, we have
\eqn\emr{N\sim r_0M,}
exactly the same relation as in the flat background. Note that this
relation has nothing to do with the detailed formula for $r_0$, it is
a result of the definition of the invariant mass and the uncertainty
relation. If $N$ indeed can be regarded as the entropy of the
black hole in the pp-wave background, using \transd\ and 
\emr, we have
\eqn\mast{r_0^{10}\sim {GMR\over \mu},\quad S\sim [{GR\over 
\mu}]^{{1\over 10}}M^{{11\over 10}},}
in contrast to the relations in the flat background
\eqn\fmast{r_0^8\sim GM,\quad S\sim G^{{1\over 8}}M^{{9\over 8}}.}

\vskip1cm

\noindent {\it Appendix}

We compute the exact Green's function of a massless scalar in the
background \metric. To simplify formulas, assume $\mu=3$, or alternatively
rescale $x^\pm$ to absorb $\mu$. The Green's function satisfies
\eqn\soug{(-\p_+\p_--h_{++}\p_-^2+\p^2)G(x,y)=\delta^{11}(x-y),}
where $h_{++}$ is the coefficient of $(dx^+)^2$ in \metric with $\mu=3$,
$\p^2$ is the Laplacian in the 9 dimensional transverse space.

Let 
\eqn\lpg{G(x^\pm-y^\pm,x,y)=\int {dk_+dk_-\over (2\pi)^2}e^{ik_+(x^+-y^+)
+ik_-(x^--y^-)}G(x,y),}
where $x,y$ are transverse coordinates, then
\eqn\trasg{(k_+k_-+h_{++}k_-^2+\p^2)G(x,y)=\delta^9(x,y).}
Apparently, the Green's function in \trasg\ can be expressed as a sum of 
harmonic eigenstates in the transverse space. For $A=i$, the
eigenstates are
\eqn\ieigen{\phi_{n_i}(x_i)=\left({\sqrt{|k_-|}\over 2^{n_i}n_i!\sqrt{\pi}}
\right)^{\half}H_{n_i}(\sqrt{|k_-|}x_i)e^{-\half |k_-|x_i^2}.}
For $A=a$, the eigenstates are
\eqn\aeigen{\phi_{n_a}(x_a)=\left({\sqrt{|k_-|}\over 2^{n_a+1}n_a!\sqrt{\pi}}
\right)^{\half}H_{n_i}(\half\sqrt{|k_-|}x_a)e^{-{1\over 4} |k_-|x_a^2}.}
The eigenstates of the operator $k_+k_-+h_{++}k_-^2+\p^2$ are
just products
\eqn\eprod{\Phi_\lambda(x)=\prod_i\phi_{n_i}(x_i)\prod_a\phi_{n_a}
(x_a).}
Finally, the Green's function of \trasg\ is given by
\eqn\trag{G(x,y)=\sum_{\{n_i,n_a\}}{1\over k_+k_- -|k_-|\sum 
(2n_i+1)+\half (2n_a+1)}\Phi_\lambda (x)\Phi_\lambda (y).}

Substitute \trag\ into \lpg\ and perform the integral over $k_+$,
we obtain a factor $\theta (x_+)$ (for the retarded Green's function)
and a product of functions depending on $n_i,x_i$ or $n_a,x_a$. 
Following \samir, we use the following identity
\eqn\suh{\sum_n{1\over n!}H_n(\sqrt{\omega}x)H_n(\sqrt{\omega}
y)({z\over 2})^n={1\over\sqrt{1-z^2}}\exp(\omega\left({2xyz-(x^2+y^2)
z^2\over 1-z^2}\right)}
to obtain a closed form of the integrand in the integral over $k_-$.
In the end, up to a numerical factor, the retarded  Green's function
can be expressed as
\eqn\retg{G(x,y)=I_+-I_-,}
where
\eqn\iret{\eqalign{I_+&=\cos^3(x^+-y^+)\int_0^\infty dk_-k_-^{7/2}
e^{ik_-(\Phi+i\epsilon)},\cr
\Phi&=2(x^--y^-)\sin 2(x^+-y^+)+2xy-(x^2+y^2)\cos 2(x^+-y^+)
\cr
&+\left(2\tilde{x}\tilde{y}-(\tilde{x}^2+\tilde{y}^2)\cos (x^+-y^+)
\right)\cos (x^+-y^+),}}
where $xy=x_ix_i$, $\tilde{x}\tilde{y}=x_ay_a$ and so on. $I_-$
is given by the same formula as \iret\ with the $k_-$ in the
exponential replaced by $-k_-$, of course the sign  of $\epsilon$
must be switched too, to guarantee the convergence of the $k_-$
integral. We can use the Green's function obtained in this appendix
to repeat the analysis in the main text.

Acknowledgments. 

This work was supported by a grant of NSC, and by a 
``Hundred People Project'' grant of Academia Sinica and an outstanding
young investigator award of NSF of China.

\vfill
\eject

\listrefs
\end